\documentclass[final,5p,times,twocolumn]{elsarticle}
\usepackage{graphicx}
\usepackage{amssymb}
\usepackage{amsmath}
\usepackage{bbm}
\usepackage{bm}
\usepackage{lscape}
\usepackage{color}
\journal{Physics Letters A}
\begin{document}
\begin{frontmatter}

\title{Explosive synchronization in coupled nonlinear oscillators on multiplex network}
\author{Umesh Kumar Verma}
\author{G. Ambika\corref{cor1}}

\address{Indian Institute of Science Education and Research(IISER) Tirupati, Tirupati, 517507, India}
\cortext[cor1]{g.ambika@iisertirupati.ac.in}

\date{\today}

\begin{abstract}
 We report the emergence of explosive synchronization in a multiplex network where oscillators on the first layer are coupled with attractive coupling and those on the second layer, coupled with repulsive coupling. With Stuart-Landau and  FitzHugh-Nagumo oscillators as the nodal dynamics, we consider non-local and mean-field intralayer couplings. We establish that explosive synchronization occurs in the multiplex network in the presence of Gaussian noise, and the transition is first order with hysteresis to a state of complete intra-layer and in-phase interlayer synchronization. The width of the hysteresis depends on the range of intralayer attractive coupling, strength of interlayer coupling and noise. We also see how to have control over this induced transition so that the explosive nature, if undesirable, can be converted to continuous type by tuning the strength of repulsive coupling or noise. 
\end{abstract}

\begin{keyword}
 Multiplex network, explosive synchronization, attractive and repulsive coupling
\end{keyword}

\end{frontmatter}

\section{Introduction}

The framework of complex networks is being extensively used to model the dynamics and transitions in many real-world complex systems in widely varying contexts.~\cite{E1}. Thus nervous systems at different organizational levels~\cite{E2}, social interaction networks with varying interactions in time~\cite{E3}, and power grid networks having different loads for connections~\cite{E4} are all examples of real complex systems that are now being studied using networks. Many of these natural and artificial systems, such as the brain, society, modern transportation networks, smart grids, etc., are made up of different constituents and/or different types of interactions. Then it is more suitable to model them using multilayer networks and, in many cases, as multiplex networks~\cite{E01,E02}. The connections between the different layers may produce various interesting phenomena, like amplitude chimera and chimera death~\cite{E03}, explosive synchronization~\cite{E04}, explosive death~\cite{E05}, etc.

The most prevalent emergent phenomenon in complex systems is synchronization, which can be at different levels of coherence in the dynamics of all connected systems. The onset of synchronization, where collective dynamics of coupled systems changes from incoherence to coherence, in most cases, is a continuous second-order phase transition~\cite{E9}. But recently reported studies indicate that an abrupt change from incoherence to coherence in the form of first-order phase transition, called explosive synchronization, is possible under specific conditions. This intriguing phenomenon of explosive synchronization(ES) was first observed by Gardenes et al. in a scale-free (SF) network of Kuramoto oscillators~\cite{E10}. Since then, ES has also been reported in a large variety of network topologies and a variety of frequency distributions of the phase oscillators~\cite{E06, E07,E08}. The analytical expression for the critical coupling for this abrupt transition has been derived using a mean-field approach in a degree-frequency correlated SF network~\cite{E11}. Subsequent studies established this transition in an adaptive complex network model where the link weights evolve with an anti-Hebbian rule and weaken the connection between pairs of coherent nodes~\cite{E12}. It is also reported in periodic oscillators on a single-layer network with attractive and repulsive interactions~\cite{E13,E23}and in  environmentally coupled systems~\cite{E14}. Explosive synchronization (ES) occurs when a small perturbation or change in parameter can lead to a discontinuous transition to synchronization on systems coupled on a network. In recent studies, this is made possible by introducing degree-frequency correlation, frequency-coupling strength correlation, inertia, or adaptively controlled phase oscillators in single layer networks~\cite{E10,E04,E012}.  

Very recently, ES in multilayer networks has received much attention and has relevance in understanding dynamical transitions in many complex systems. Since the first report of explosive transition to synchronization in a multilayer network of phase oscillators ~\cite{E15,E16}, a few studies of ES on multiplex networks illustrate the additional flexibility and tunability that multiplexing provides in inducing ES. In multiplex networks, ES is achieved by introducing phase shift in interlayer coupling terms~\cite{E01}, frequency mismatch~\cite{E09}, time-delayed intra-layer coupling~\cite{E010}, phase-frustration~\cite{E011}, inhibitory coupling~\cite{E012}, random pinning of inter layer interactions~\cite{E013}, inter layer adaptation~\cite{E014} etc between the two layers of the multiplex network.

At present, all these studies of ES are on networked Kuramoto phase oscillators. To the best of our knowledge, the transition to ES on coupled nonlinear oscillators on multiplex networks remains unexplored. While phase oscillators can model phase relationships in many real-world systems, variations or transitions in the amplitude of the oscillations are also relevant in many situations. In this work, we consider two such oscillators, Stuart-Landau (SL) and FitzHugh-Nagumo (FHN), as model systems and study the possibility of inducing ES in them when coupled on a bilayer multiplex framework.   

Our understanding till now is that continuous transition to synchronization in a network is initiated by the formation of a large cluster by merging of small synchronization clusters~\cite{E015,E016}. This then grows gradually, adding smaller local clusters to it. However, if the growth of the large cluster is suppressed by introducing asymmetry or any other types of perturbation or frustration, there will be only multiple, small-sized, synchronized clusters. But when the critical threshold is reached, the sudden unification of all the clusters happens, resulting in ES. In the present study, we achieve this by having attractive and repulsive nonlocal coupling in the two layers with ring topology. We take two types of coupling, nonlocal diffusive and mean-field couplings, and find that in the presence of noise, the two layers undergo ES with hysteresis simultaneously.  

For this study, we consider a multiplex network with periodic oscillators where first layer oscillators are coupled with attractive coupling and second layer oscillators are coupled with repulsive coupling. The heterogeneous nature of couplings of attractive and repulsive types is relevant in various contexts like neuronal networks, where various spatio-temporal patterns are induced through excitatory and inhibitory synapses~\cite{E17,E18}. Thus the mixed-coupling of attractive-repulsive type results in amplitude death in the relay system~\cite{N1}, while the effect of additional repulsive links can induce oscillation death in a network of globally coupled oscillators~\cite{N2}. In the multiplex architecture, intralayer synchronization coexists with antiphase dynamics between coupled systems of different layers in the presence of attractive intralayer and repulsive interlayer connections~\cite{N3}. The various emergent phenomenon such as chimera states, solitary states, extreme events, antiphase synchrony, amplitude or oscillation death,  cluster states, and traveling waves can appear in ensembles of oscillatory units under the simultaneous action of attractive and repulsive interactions~\cite{N4}.

While continuous transitions to synchronization are gradual and easy to control, explosive transitions may occur suddenly and can have undesirable effects. Thus an abrupt state of transition is proposed as a potential mechanism of the hypersensitive Fibromyalgia brain, which relates to an optimally balanced state~\cite{E017}. We note that hypersensitive responses to external stimuli are observed in other systems and can be related to cascading failures in power grids, transitions in an electronic circuit, loss and recovery of consciousness in general anesthesia, and epileptic seizures in the brain~\cite{E018,E9,E019} etc. So converting explosive to continuous transition is also required as a control mechanism. In this context, our study shows that by increasing the range of nonlocal interactions, or inter-layer coupling, or strength of noise, we can reduce the hypersensitivity and thus control ES.

The paper is organized as follows: In Sec.2, the multiplex network with attractive and repulsive coupling is presented with the relevant order parameters. In Sec.3, we discuss our main results for nonlocal diffusive coupling among SL and FHN oscillators, and in Sec.4, the same analysis but with a mean-field coupling scheme. We summarize our main results and conclusions in Sec.5.

\section{Model system and methodology}
We construct a multiplex network with N nonlinear oscillators on each layer, with the oscillators on first layer (L1) coupled with attractive coupling, and those on second layer (L2) coupled with repulsive coupling. The dynamics of such a model system in general is thus given by,

\begin{eqnarray}
\dot{\bm{X}}_{i,1} & = \bm{F}(\bm{X}_{i,1})+\lambda_a \sum_{j=1}^{N} G_1(x^1_{j,1},x^1_{i,1})+\epsilon x^1_{i,2}+\gamma\xi_i(t)\nonumber\\ 
 \dot{\bm{X}}_{i,2} &= \bm{F}(\bm{X}_{i,2})-\lambda_r \sum_{j=1}^{N}G_2(x^1_{j,2},x^1_{i,2})+\epsilon x^1_{i,1}+\gamma\xi_i(t)\nonumber\\ 
\label{eq1}
\end{eqnarray}
where $i=1,2,\ldots,N$. Here $\bm{X}_{i,l}= \begin{bmatrix} x^1_{i,l} & x^2_{i,l} & \ldots & x^m_{i,l} \end{bmatrix}^{T}$ represent $m$--dimensional state space of the oscillator whose intrinsic dynamics is given by 
$\bm{F}(\bm{X}_{i,l})=\begin{bmatrix} f^1_{i,l}(\bm{X}_{i,l}) &f^2_{i,l}(\bm{X}_{i,l})& \ldots & f^m_{i,l}(\bm{X}_{i,l}) \end{bmatrix}^{T}$, 

where $l=1,2$ represent the two layers of the multiplex network. $G_1$ and $ G_2$ are coupling matrices having dimensions $N \times N$  and their elements depend on the specific coupling scheme chosen. Oscillators in the first layer L1 interact with oscillators in the second layer L2 with multiplex like $i$ to $i$ coupling of feedback type of strength $\epsilon$. $\lambda_a$, and $\lambda_r$  represent the strengths of attractive and repulsive coupling of the oscillators on L1 and L2, respectively. Additionally, we include Gaussian noise $ \xi_i$ with strength $\gamma$ on both layers. We consider two typical nonlinear oscillators, Stuart- Landau (SL) and FitzHugh-Nagumo (FHN), and two types of coupling schemes as nonlocal diffusive and mean-field types.

In order to characterize the collective behaviour of the coupled systems, the synchronization order parameter usually considered is derived from ~\cite{E24},

\begin{equation}
\begin{split}
r(t)e^{i\psi (t)}=\frac{1}{N}\sum_{j=1}^{N}e^{i\theta_j(t)}
\end{split}
\end{equation}

where $\psi(t)=\frac{\sum \theta_j}{N}$ is the average phase of oscillators at time t and $r(t)$ is the local order parameter at each instant. For nonlinear oscillators like LS or FHN with variables $x$ and $y$, instantaneous phase $\theta_i$ of each oscillator is calculated by $\tan^{-1}(y_i/x_i)$. The global order parameter $R = |\bar{r}(t)|$ lies in the interval $R \in [ 0 , 1 ]$, measures the phase coherence of the system. For $R\approx 1$ system show phase synchronization, while for $R\approx 0$ system show completely incoherent state. It is clear that this order parameter indicates only the phase coherence among the oscillators. Hence to check the completely synchronized state, we define another order parameter, which is the average synchronization error, $ S$, where, 
\begin{equation}
{ S = \left\langle \sqrt{\frac{1}{N} \sum_{i=1}^N \left[(x_i- \bar{x})^2\right]} \right\rangle.} 
\label{eq:eq3}
\end{equation}
Here, $<..>$ represents the long-time average. Clearly, for complete synchronization $S=0$ while for an asynchronous state $S>0$ . 

To study the nature of the transition in the system, we follow the adiabatic method of tuning the parameters. For forward continuation, first $R(0)$ or $S(0)$ is calculated for some random initial conditions and then the value of $\lambda_a$ is gradually increased from $\lambda_a(0)$ to $\lambda_a(max)$, in steps of size $\delta\lambda_a = 0.01$. The final state of the previous $\lambda_a$ is used as the initial condition for the next $\lambda_a$. The backward continuation is also performed in the same way by gradually decreasing the value of $\lambda_a$ from $\lambda_a(max)$ to $\lambda_a(0)$.

\begin{figure}
\includegraphics[width=0.5\textwidth]{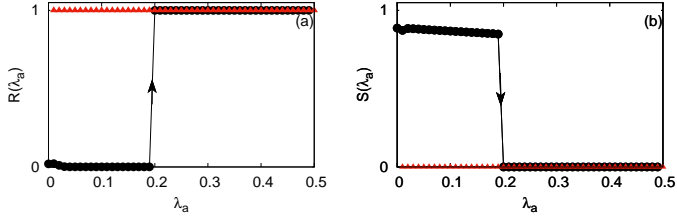}
\caption{ Transition to synchronization in multiplex network under attractive and repulsive couplings in the absence of noise. (a) Order parameter (R) and (b) synchronization error (S)  as a function of coupling strength $\lambda_a$ with forward (circle)  and backward  continuation (triangle) for first layer of coupled SL oscillators at $p_1=30$.  The other parameters are $p_2=25$, $\epsilon=1$, $\gamma=0.00$, $\lambda_r=0.1$ and $N=100$.}
\label{fig:f1}
\end{figure}
\section{Oscillators under attractive  and repulsive nonlocal coupling}

We first consider non-identical SL oscillators as nodal dynamics on the multiplex network, with L1, having attractive nonlocal coupling, and L2, repulsive nonlocal coupling in ring topology. The dynamics of the multiplex network as from Eq.~\ref{eq1} is given by,

\begin{eqnarray}
\dot{x}_{i,1}&=&(1-x_{i,1}^2-y_{i,1}^2)x_{i,1}-\omega_1(i) y_{i,1}\nonumber\\
&&+ \frac{\lambda_a}{2p_1}\sum_{j=i-p_1}^{i+p_1}(x_{j,1}-x_{i,1})+\epsilon x_{i,2}+\gamma\xi_i(t) \nonumber\\
\dot{y}_{i,1}&=& (1-x_{i,1}^2-y_{i,1}^2)y_{i,1}+\omega_1(i) x_{i,1}+\gamma\xi_i(t) \nonumber\\
\dot{x}_{i,2}&=&(1-x_{i,2}^2-y_{i,2}^2)x_{i,2}-\omega_2(i) y_{i,2}\nonumber\\
&&- \frac{\lambda_r}{2p_2}\sum_{j=i-p_2}^{i+p_2}(x_{j,2}-x_{i,2})+\epsilon x_{i,1}+\gamma\xi_i(t) \nonumber\\
\dot{y}_{i,2}&=& (1-x_{i,1}^2-y_{i,1}^2)y_{i,1}+\omega_2(i) x_{i,2}+\gamma\xi_i(t) \nonumber\\
\label{sl:eq0}
\end{eqnarray}

where $i = 1, 2, . . . ,N$. $x_i$ and $y_i$ are the state variables of the $i^{th}$ SL oscillator. The frequency of SL oscillators  $\omega_1(i)$ and $\omega_2(i)$ are chosen with Gaussian distribution in $(1, 2)$.  The interaction between the oscillators in L1 is controlled by $\lambda_a$ and $p_1$, whereas the interaction in L2 is controlled by $\lambda_r$ and $p_2$. $p_1$ and $p_2 \in {(1,N/2)}$ corresponds to the number of coupled neighbors in each direction on L1 and L2 respectively. For local coupling $p_i = 1$, for global coupling $p_i = N$ and for non-local coupling value of $p_i$ is $(1 < p_i < N/2)$, (where $i=1,2$). The layers are multiplexed by feedback coupling of strength $\epsilon$ and evolve under added Gaussian noise of strength $\gamma$.

\begin{figure}
\includegraphics[width=0.5\textwidth]{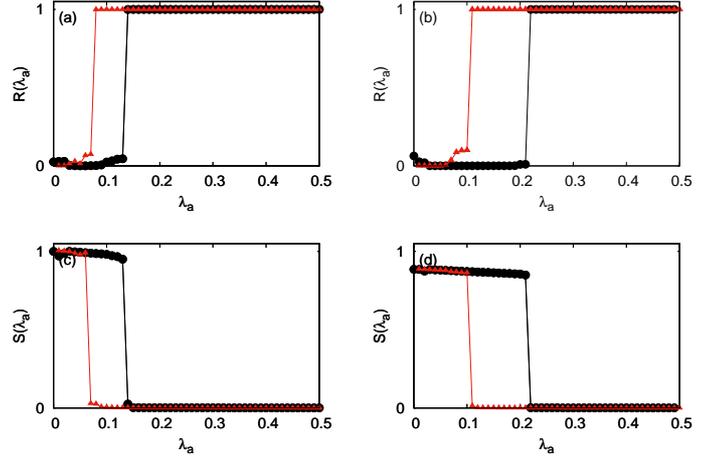}
\caption{ Explosive transition to synchronization with hysteresis in multiplex network with attractive  and repulsive nonlocal coupling in the presence of noise: Order parameter (R) and synchronization error (S) as a function of coupling strength $\lambda_a$ with forward (circle) and backward  continuation (triangle) in coupled SL oscillators on L1 with gaussian noise strength $\gamma=0.001$:  (a,c)$p_1=25$, and (b,d)$p_1=40$. Here $p_2=25$, $\epsilon=1$, $\lambda_r=0.1$, and $N=100$.  The nature of transition and transition point are exactly the same for the second layer.}
\label{fig:f2}
\end{figure}

First, we study the dynamics of the above multiplex network in the absence of noise (i.e. $\gamma=0$). We set $p_1=30$,$p_2=25$, $\lambda_r=0.1$ and calculate the order parameter $R(\lambda_a)$ and $S(\lambda_a)$ adiabatically in both forward and backward directions for $N=100$ oscillators in each layer. In the forward direction, both order parameters show a sudden transition, which refers to an abrupt transition from an asynchronous state to a complete synchronized state as shown in Fig.~\ref{fig:f1}(a) and (b), respectively. In the case of backward transition, we observe no transition in the order parameters. 

Next, we add a small strength of Gaussian noise in both layers, with $\gamma=0.001$, and calculate order parameter $R(\lambda_a)$ and $S(\lambda_a)$ adiabatically in both forward and backward directions, keeping all other parameters the same. Their variations with $\lambda_a$ are shown in Fig.~\ref{fig:f2}(a) and (c) respectively.  We observe that in the presence of noise, the forward and backward transitions occur at different points showing hysteresis, where both completely synchronized and asynchronous states co-exist. We note that hysteresis and backward transition point are observed only in the presence of noise.  Due to the presence of noise, in the backward transition, the initial conditions in each value of coupling strength can be slightly different, for all the synchronized oscillators. Then at a critical coupling strength, the system goes to desynchronized state.

We also note that the nature of transition and forward and backward transition points are the same for both layers. When we increase $p_1$ from $p_1=25$ to $p_1=40$, we observe, width of the hysteresis is increased, as shown in Fig. ~\ref{fig:f2}(b) and (d). 

\begin{figure}
\includegraphics[width=0.5\textwidth]{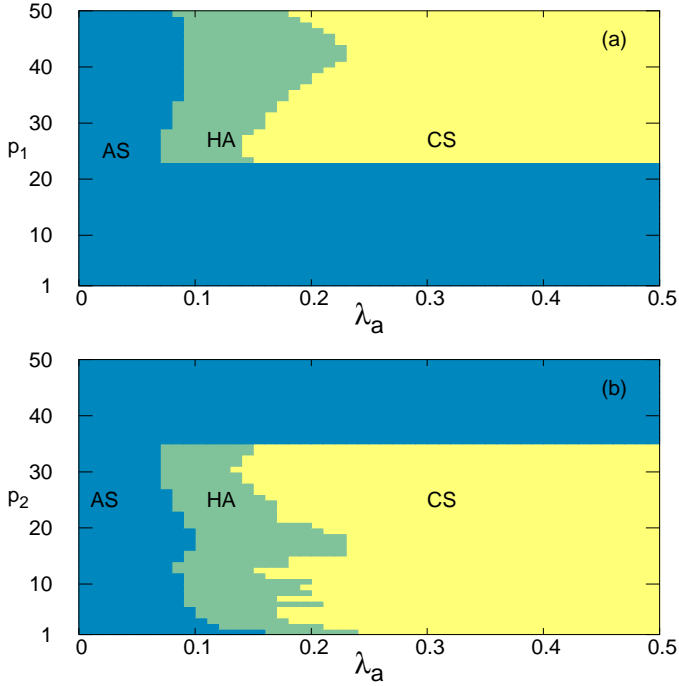}
\caption{ Phase diagrams indicating the different dynamical regimes for multiplex network with attractive and repulsive nonlocal couplings for  first layer. (a) $(\lambda_a-p_1)$ at $p_2=30$ and (b) $(\lambda_a-p_2)$ at $p_1=30$ .  Here three different regimes, AS, CS, and HA, indicate asynchronous state, completely synchronized state, and hysteresis area (where both asynchronous and synchronous state coexists). The other parameters are set at $\gamma=0.001$ and $\lambda_r=0.1$,  $p_2=30$, $\epsilon=1$, and $N=100$. The phase diagram for the second layer also has similar pattern and transition points are exactly the same.}
\label{fig:f3}
\end{figure}

\begin{figure}
\includegraphics[width=0.5\textwidth]{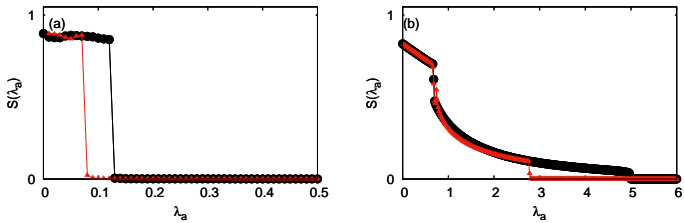}
\caption{ Transition from explosive to continuous synchronization as $\lambda_r$ is increased. The variation of $S$ as a function of coupling strength $\lambda_a$ with forward (circle) and backward continuation (triangle) of coupled SL oscillators: (a)$\lambda_r=0.1$, (b)$\lambda_r=0.8$. We set $p_1=25$, $p_2=25$, $\epsilon=1$, $\gamma=0.001$ and $N=100$.}
\label{fig:f4}
\end{figure}

We scan the parameter plane of ($\lambda_a, p_1$)and plot the phase diagram of coupled SL oscillators, indicating the different dynamical regimes of the first layer's oscillators, which is shown in Fig.~\ref{fig:f3}(a). Here, AS, CS, and HA represent the asynchronous regime,  completely synchronized regime, and hysteresis area, respectively. From this diagram, we can see that transition from an asynchronous state to a complete synchronous state occurs for a value of $p_1>22$ and the hysteresis area increases with increasing the value of $p_1$ with a maximum at $p_1= 42$. The parameter space for $(p_2-\lambda_a)$ at  $p_1=30$ is plotted in Fig.~\ref{fig:f3}(b). We can see that the transition occurs only for $p_2<35$ with HA maximum for $p_2<15$.

For $p_1=25$ and $p_2=25$, we study the role of repulsive coupling on the nature of transitions. We find $\lambda_r=0.1$ the order parameter $S$, shows discontinuous transition (Fig.~\ref{fig:f4}(a)). As the strength of repulsive coupling is increased to $\lambda_r=0.8$, the transition becomes continuous, as is clear from Fig.~\ref{fig:f4}(b). Thus by changing $\lambda_r$, we can change the nature of transition from abrupt to continuous.

\begin{figure}
\includegraphics[width=0.5\textwidth]{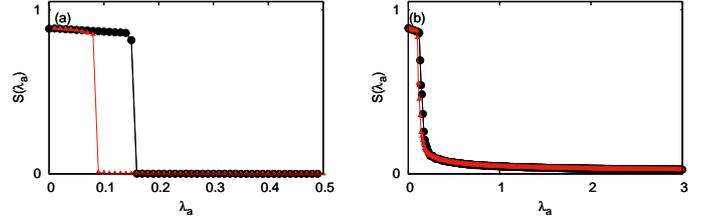}
\caption{ Transition from explosive to continuous synchronization with increasing noise strength. Order parameter S for synchronization as a function of coupling strength $\lambda_a$ with forward (circle)  and backward  continuation (triangle) for coupled SL oscillator: (a)$\gamma_=0.001$, (b)$\gamma=0.15$. Here $p_1=30$, $p_2=25$, $\epsilon=1$, $\lambda_r=0.1$, and $N=100$.}
\label{fig:f5}
\end{figure}

We also investigate the effect of noise on the nature of transition in the multiplex network. With strength of noise $\gamma=0.001$ for $p_1=30$, and $p_2=25$, we observe a sudden transition with hysteresis, which is shown in  Fig.~\ref{fig:f5}(a). When we increase the strength of noise $\gamma$, we observe the width of hysteresis  decreasing. Further, when $\gamma>0.1$ the nature of transition is changed from discontinuous to continuous as shown Fig.~\ref{fig:f5}(b)

\begin{figure}
\includegraphics[width=0.5\textwidth]{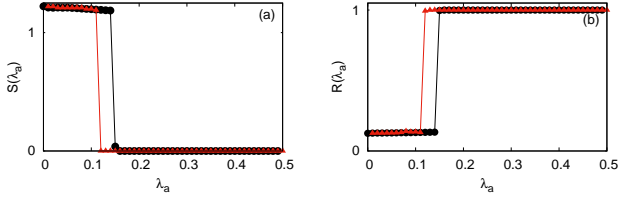}
\caption{Explosive synchronization in coupled FHN oscillators on multiplex network with attractive and repulsive nonlocal coupling: (a) Order parameter S  and (b) R for synchronization  as a function of coupling strength $\lambda_a$ with forward (circle) and backward  continuation (triangle)  at $p_1=40$. The other parameters are set at $\lambda_r=0.1$, $p_2=30$, $\epsilon=1$, $\gamma=0.001$, and $N=100$.}
\label{fig:f6}
\end{figure}

We extend our study to another nonlinear system, forming a multiplex network with N FitzHugh-Nagumo (FHN) oscillators on each layer. The dynamics of multiplex network, in this case is given by,

\begin{eqnarray}
\tau \dot{x}_{i,1}&=&x_{i,1}-\frac{x_{i,1}^3}{3}-y_{i,1}+ \frac{\lambda_a}{2p_1}\sum_{j=i-p_1}^{i+p_1}(x_{j,1}-x_{i,1})\nonumber\\
&&+\epsilon x_{i,2}+\gamma\xi_i(t) \nonumber\\
\dot{y}_{i,1}&=& x_{i,1}+\beta+\gamma\xi_i(t) \nonumber\\
\tau \dot{x}_{i,2}&=&x_{i,2}-\frac{x_{i,2}^3}{3}-y_{i,2}- \frac{\lambda_r}{2p_2}\sum_{j=i-p_2}^{i+p_2}(x_{j,2}-x_{i,2})\nonumber\\
&&+\epsilon x_{i,1}+\gamma\xi_i(t) \nonumber\\
\dot{y}_{i,2}&=& x_{i,2}+\beta+\gamma\xi_i(t) \nonumber\\
\label{Eq.fhn0}
\end{eqnarray}

where $i=1,2,.....N$, $\beta=0.5$, and $\tau=0.05$. The other parameters are as given in Eq.~\ref{sl:eq0}.
We set $p_2=30$, $\lambda_r=0.1$ and calculate the order parameters $S(\lambda_a)$ and $R(\lambda_a)$ adiabatically in both forward and backward directions, which is shown in Fig.~\ref{fig:f6} (a,b) for $p_2=40$ respectively. Here we observe a sudden transition with hysteresis. By changing the parameters, we get  qualitatively similar results like those with SL system presented above.

\begin{figure}
\includegraphics[width=0.5\textwidth]{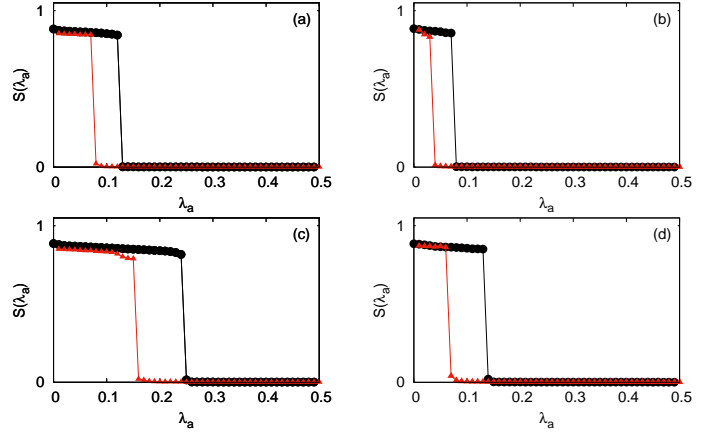}
\caption{Explosive synchroniation in coupled SL oscillators on multiplex network with attractive and repulsive mean-field coupling: Forward (circle) and backward continuation (triangle) as a function of coupling strength  $\lambda_a$ for first layer:  (a) $Q_1=1.0$, $Q_2=1.0$. (b)$Q_1=1.0$, $Q_2=0.5$ (c)$Q_1=0.5$, $Q_2=1.0$ and (d) $Q_1=0.5$, $Q_2=0.5$.  The other parameters are $\lambda_r=0.1$, $\epsilon=1$, $\gamma=0.001$, and $N=100$.}
\label{fig:f7}
\end{figure}

\begin{figure}
\includegraphics[width=0.5\textwidth]{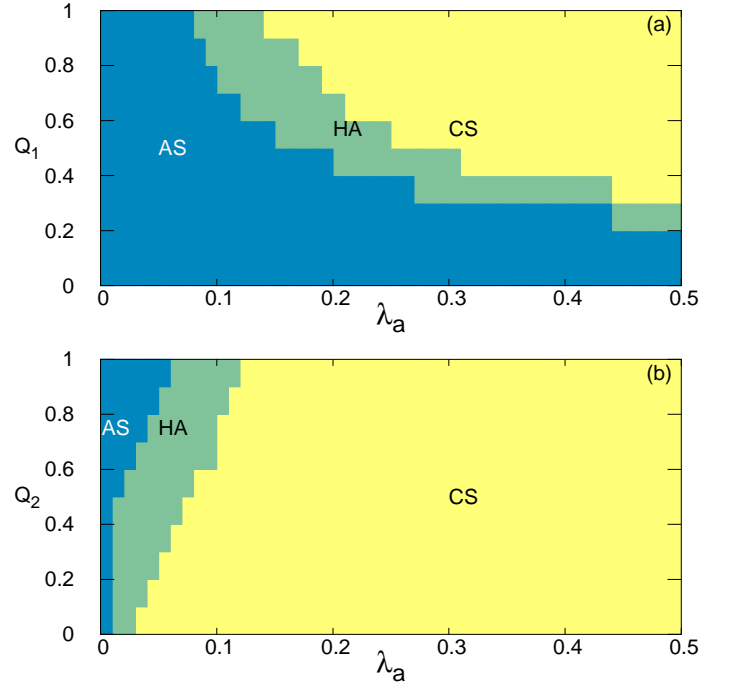}
\caption{Phase diagrams indicating the different dynamical regimes of first layer of SL oscillators coupled with mean-field attractive and repulsive coupling in the parameter plane: $(\lambda_a-Q_1)$  for $Q_2=1.0$, and (b)$(\lambda_a-Q_2)$ for $Q_1=1.0$.   The other parameters are $\gamma=0.001$, $\lambda_r=0.1$, $\epsilon=1$, and $N=100$.}
\label{fig:f8}
\end{figure}


\section{ Oscillators under attractive and repulsive mean-field coupling}

Now we consider a multiplex network where intralayer coupling is of the mean-field type.  The N non-identical Stuart-Landau oscillators (SL) on each layer, with this type of attractive and repulsive coupling evolve as given by

\begin{eqnarray}
\dot{x}_{i,1}&=&(1-x_{i,1}^2-y_{i,1}^2)x_{i,1}-\omega_1(i) y_{i,1}\nonumber\\
&&+\lambda_a (Q_1\bar{x}_{i,1}-x_{i,1})+\epsilon x_{i,2}+\gamma\xi_i(t) \nonumber\\
\dot{y}_{i,1}&=& (1-x_{i,1}^2-y_{i,1}^2)y_{i,1}+\omega_1(i) x_{i,1}+\gamma\xi_i(t) \nonumber\\
\dot{x}_{i,2}&=&(1-x_{i,2}^2-y_{i,2}^2)x_{i,2}-\omega_2(i) y_{i,2}\nonumber\\
&&- \lambda_r(Q_2\bar{x}_{j,2}-x_{i,2})+\epsilon x_{i,1}+\gamma\xi_i(t) \nonumber\\
\dot{y}_{i,2}&=& (1-x_{i,1}^2-y_{i,1}^2)y_{i,1}+\omega_2(i) x_{i,2}+\gamma\xi_i(t) \nonumber\\
\end{eqnarray}
where $i=1,2,....N$ and $Q_1$, $Q_2$ with $0\leq Q_j<1 $ (where $j=1,2$) is the intensity of the mean-field. The other parameters are as mentioned in the previous section in Eqn.~\ref{sl:eq0}.

First we set $Q_1=1$, $Q_2=1$ and $\lambda_a=0.1$, and compute the order parameter $S(\lambda_a)$, in both forward and backward directions and the results are shown in Fig.~\ref{fig:f7}(a). In this case the transition is sudden with hysteresis which show the emergence of ES in the network. As we decrease the value of $Q_1$ from $Q_1=1$ to $Q_1=0.5$ , we observe that both forward and backward transition points are shifted to lower values, which is shown in Fig.~\ref{fig:f7}(b). This  indicates that ES can occur for a smaller value of $\lambda_a$, by decreasing $Q_1$. Also decreasing the value of $Q_1$  to $Q_1=0.5$, the width of hysteresis region increases (Fig.~\ref{fig:f7}(c)and (d)). 

The effect of changing other parameters is clear from the parameter space $(\lambda_a-Q_1)$ and $(\lambda_a-Q_2)$ in Fig.~\ref{fig:f8}(a) and (b) respectively. Fig.~\ref{fig:f8}(a) indicates that hysteresis and transition from asynchronous state to complete synchronized state are observed only for $Q_1>0.2$, while as in Fig.~\ref{fig:f8}(b) hysteresis and transition can be observed for the range of values of $Q_2$ as shown. In this case also we can change the nature of transition from explosive to continuous synchronization by increasing the repulsive coupling strength $\lambda_r$ or increasing the strength of noise $\gamma$.

\begin{figure}
\includegraphics[width=0.5\textwidth]{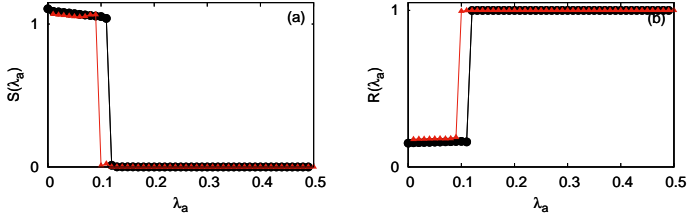}
\caption{Explosive synchronization in coupled FHN oscillators on multiplex network with attractive and repulsive mean-field coupling: (a) S and (b) R for forward (circle) and backward contiuation (triangle) as a function of coupling strength $\lambda_a$ at $Q_1=1$ and $Q_2=1$. We set $\lambda_r=0.1$, $\epsilon=1$, $\gamma=0.001$, and $N=100$.}
\label{fig:f9}
\end{figure}

For the multiplex network with FHN oscillators, on each layer, coupled with attractive and repulsive mean-field coupling, the dynamics is given by,

\begin{eqnarray}
\tau \dot{x}_{i,1}&=&x_{i,1}-\frac{x_{i,1}^3}{3}-y_{i,1}+ \lambda_a (Q_1\bar{x}_{i,1}-x_{i,1})\nonumber\\
&&+\epsilon x_{i,2}+\gamma\xi_i(t) \nonumber\\
\dot{y}_{i,1}&=& x_{i,1}+\beta+\gamma\xi_i(t) \nonumber\\
\tau \dot{x}_{i,2}&=&x_{i,2}-\frac{x_{i,2}^3}{3}-y_{i,2}- \lambda_r (Q_2\bar{x}_{i,2}-x_{i,2})\nonumber\\
&&+\epsilon x_{i,1}+\gamma\xi_i(t) \nonumber\\
\dot{y}_{i,2}&=& x_{i,2}+\beta+\gamma\xi_i(t) \nonumber\\
\end{eqnarray}

Here all parameters are the same in Eq.~\ref{Eq.fhn0}. In Fig.~\ref{fig:f9}(a,b), the order parameters $S$ and $R$ for $Q_1=1$ and  $Q_2=1$ indicate a sudden transition in both forward and backward directions establishing ES in the system.


\section{Conclusion}

In this letter, we present the study on the nature of transitions in the emergent dynamics on a multiplex network of coupled nonlinear oscillators. The sensitive state required for explosive synchronization is generated in the network by balancing the couplings on both the layers with an attractive coupling on the first layer and repulsive on the second layer and multiplexing them with a feedback type of inter-layer coupling.
We report that for chosen values of the strength of couplings and range of couplings, the network can support explosive synchronization. Moreover, hysteresis in the transition is observed with added noise of required strength. We illustrate this phenomenon for two types of nonlinear oscillators, Stuart-Landau and FHN oscillators, as nodal dynamics and intra-layer couplings of diffusive and mean-field types. We get qualitatively similar results in these cases, which establishes the generality of the scheme.

Our study indicates that explosive synchronization is induced in the presence of Gaussian noise for a higher value of attractive nonlocal connections compared to that of repulsive nonlocal connections. The width of the hysteresis in this case, crucially depends on the range of intralayer nonlocal couplings and the strength of the noise. 

In the multiplex network with mean-field attractive-repulsive coupling, the transition point is dependent on the intensity of the mean-field interactions. However, the width of hysteresis depends on the mean-field attractive coupling and noise.

We observe that for both types of couplings while each layer is in complete synchroniation, the interlayer synchroniztion between them is  in-phase only. The explosive nature of transition also depends on the interlayer coupling strength $\epsilon$. The transition is continuous in the system for a higher value of coupling strength $\epsilon>1.2$ for the parameter values chosen.

As reported, the explosive nature of the transition may not be desirable in many real-world systems. Our study indicates that the sensitive, balanced state required for ES can be altered, and the nature of transition changed to continuous type by tuning the intralayer and interlayer couplings. While a small amount of noise is required for the transition, increasing noise strength decreases the width of hysteresis and can lead to continuous transition. 

\section*{Acknowledgement}

UKV acknowledges Indian Institute of Science Education and Research(IISER) Tirupati, Tirupati, India for financial support through post-doctoral fellowship.


\end{document}